# A MULTICATEGORY JET IMAGE CLASSIFICATION FRAMEWORK USING DEEP NEURAL NETWORK

Jairo José Orozco Sandoval, Vidya Manian and Sudhir Malik
Department of Electrical and Computer Engineering
Department of Physics
University of Puerto Rico, Mayaguez, PR 00681-9000

Jet point cloud images are high dimensional data structures that needs to be transformed to a separable feature space for machine learning algorithms to distinguish them with simple decision boundaries. In this article, the authors focus on jet category separability by particle and jet feature extraction, resulting in more efficient training of a simple deep neural network, resulting in a computational efficient interpretable model for jet classification. The methodology is tested with three to five categories of jets from the JetNet benchmark jet tagging dataset, resulting in comparable performance to particle flow network. This work demonstrates that high dimensional datasets represented in separable latent spaces lead to simpler architectures for jet classification.

**Introduction**
The Large Hadron Collider (LHC) situated in CERN, Switzerland, produces proton-proton (pp) collision data at the rate of 40 MHz, being the world's largest collider and producer of pp collision events [1]. With the upcoming High Luminosity (HL) experiment, the amount of particle data generated per second will reach approximately 50 TB [2]. The events at the LHC involve particle collisions, decay, and splitting. Jets are reconstructed objects composed of a collimated stream of particles that deposit energy in the Calorimeters. Jets are produced by the hadronization of quarks/gluons, which occur in the vast majority of inelastic pp collisions at the LHC. Machine Learning (ML) is utilized at all stages of the collider, including particle collision simulation, fragmentation, hadronization, and jet reconstruction [3][4][5]. Collision simulation methods include Monte-Carlo algorithms [6] and Invertible Neural Networks (INN) [7][8]. Jet reconstruction occurs at three levels: reconstruction of particle tracks from detector hits, reconstruction of vertices from tracks, and reconstruction of jet observables. ML algorithms are used in each of these categories. Specific problems in jet physics that apply ML include jet classification, flavor jet tagging, anomaly detection, pile-up mitigation, and event reconstruction.

Jets are the most ubiquitous events at the LHC. A jet is a stream of particles deposited on the detector. Four types of detectors are located at the LHC: Compact Muon Solenoid (CMS), ATLAS, ALICE, and LHCb. These detectors record the speed, mass, and charge of particles. Jet tagging is the task by which physicists identify the source particle of the jets. Jet events at the LHC are analyzed using ML methods to extract physics information, understand underlying physical processes, and search for new physics. These ML methods include the Cambridge-Aachen and Anti-kT algorithms that group multiple particles in high-energy collisions. Jet substructure analysis helps understand the subjettiness of jets. Unfolding methods are used to unfold jet observables or features from simulation and generate measurements at the detector. Flavor tagging is used to identify jets as originating from different types of quarks, such as bottom or top quarks.

The representation of jets, composed of particles, plays a crucial role in the efficacy of ML algorithms used for jet tagging. Each particle in a jet is represented by four-momentum vectors. Jets composed of these particles are also represented by four-momentum jet features, including relative eta (η), relative phi (φ), jet mass (m), and transverse momentum (Pt), which is the most basic method of representing jets. This representation allows for the calculation of invariant mass and other kinematic variables. In the calorimeter, jets are described by the energy deposits in the cells of the detector. The energy and position of the jets in the detector cell are used to reconstruct them. A Convolutional Neural Network-based architecture called PFJet-SSD is proposed in [7] for jet clustering and classification. Recently, representing jets as point clouds in a dimensional grid has gained popularity, allowing researchers to apply computer vision models and two-dimensional Convolutional Neural Networks (CNNs) to be trained on jet images and classify them.

Jet classification has become a challenging task due to the similarity among jets. In this work, we propose a novel artificial intelligence model that incorporates advanced feature extraction techniques and a deep neural network architecture to address this challenge. The objective is to enhance the accuracy and efficiency of jet classification by leveraging the rich information embedded in the jet features. Given the complexity and high-dimensional nature of the data generated at the LHC, this task is particularly demanding. However, the proposed method holds the potential to significantly improve our understanding of jet physics and contribute to the discovery of new physics phenomena by providing more precise and reliable classifications of jet events. The contributions of this research are the following:

1. **Develop an architecture for jet feature extraction:**
   - Design and implement an advanced machine learning architecture that enables precise and efficient extraction of features from jets.
   - Optimize the representation of jets through four-momentum vectors (features) and other advanced representation techniques to capture as much information as possible about the jets' properties.
2. **Apply a Deep Neural Network (DNN) for jet classification:**
   - Implement a deep neural network that utilizes the extracted features of the jets to classify them.
   - Evaluate and optimize the performance of the DNN in terms of accuracy, efficiency, and its ability to distinguish between different types of jets, overcoming the inherent challenges of jet similarity.

These goals aim to enhance the accuracy and efficiency of jet classification, significantly contributing to the analysis of LHC data and future analysis on interpretability of the AI model.

**Methodology**
The data used for developing this work is sourced from JetNet [12], a Python library specifically designed to enhance accessibility and reproducibility in machine learning (ML) research within the field of high energy physics (HEP), with a particular focus on particle jets. JetNet, built on the widely-used PyTorch framework, offers user-friendly and standardized interfaces for a variety of HEP datasets. This dataset includes jets from Gluons, light quarks, top quarks, W bosons, and Z bosons. Each jet is characterized by particle features (etarel, phirel, petrel, and mass) and jet features (type, pt, eta, and mass). By leveraging this comprehensive dataset, the study aims to accurately classify and analyze the different types of jets, thereby contributing to advancements in HEP research.

First, features are extracted from both particles and jets. For particle feature extraction, we propose a particle feature embedding based on a point cloud transformer, utilizing multihead attention and a feedforward network. This approach allows for capturing intricate relationships and dependencies between particles within a jet. For jet feature extraction, a 1D convolutional neural network (CNN) based autoencoder is used.

Once the features of both particles and jets have been obtained, two additional features—particle density and average intensity are added to enrich the dataset. With these comprehensive features, we proceed to implement the classification method using a deep neural network (DNN). The workflow of this process is illustrated in Figure 2, demonstrating the steps from feature extraction to the final classification of jets. This approach aims to enhance the accuracy and reliability of jet classification, in a computationally efficient manner.

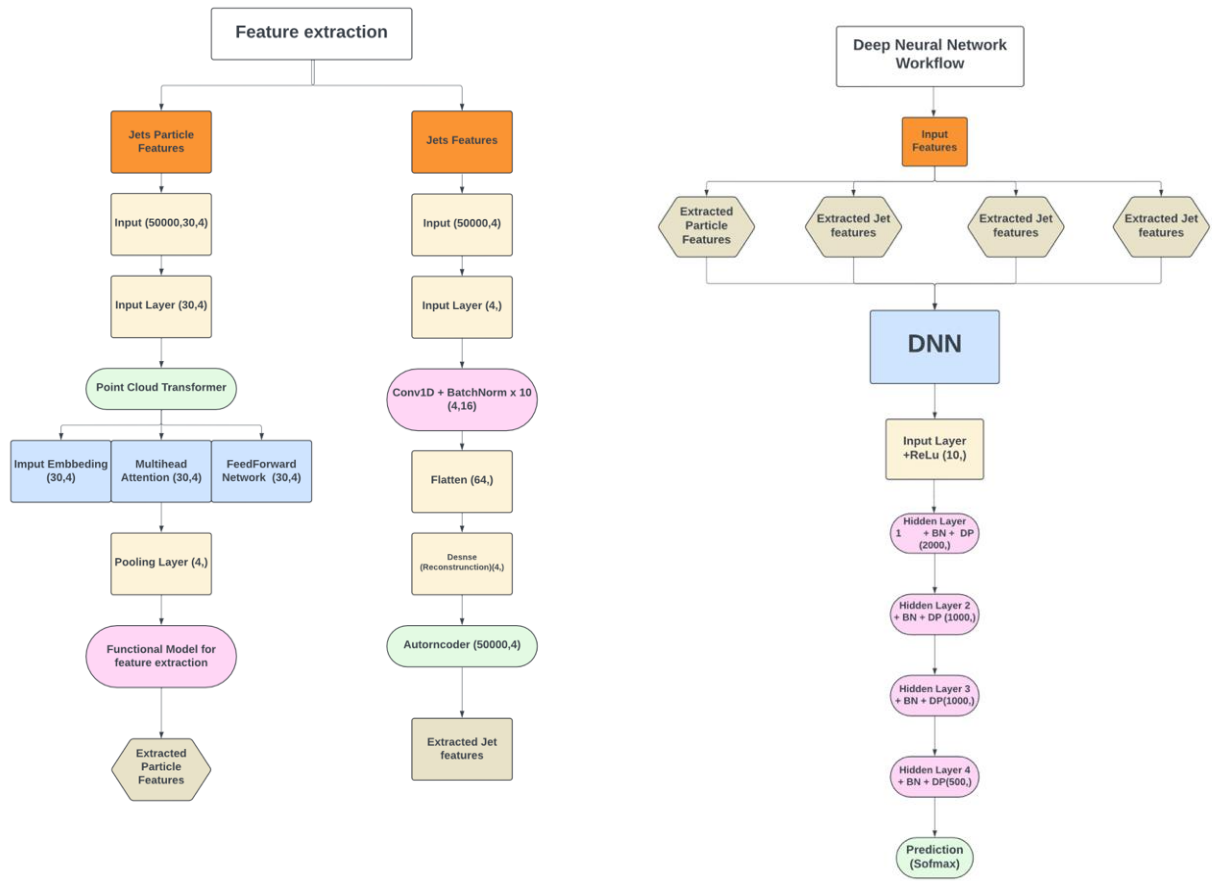

Fig 1. a) Feature extraction workflow.  b) Deep Neural Network workflow

**Preliminary Results**

The classification results for five classes abbreviated as follows: class 1: "t" (top quark), class 2: "w" (W boson) class 3: "z" (Z boson) jets, class 4: "g" (gluon), and class 5: "q" (light quark), are presented. Figure 2a shows a 3D plot of the five classes, illustrating that all classes are initially grouped together, making them difficult to distinguish. In Figure 2b, the data is shown after applying the proposed feature extraction method, revealing better grouping and distinction of the classes.

For each class, 50,000 jets were used, resulting in a total of 250,000 jets. To train the DNN 80% of the dataset was used and 20% to test the AI model. After extracting the features, the classification results are presented in the confusion matrix shown in Figure 3. The matrix indicates that the first three classes achieved classification accuracies over 80%. However, the last two classes, corresponding to gluons and light quarks, achieved accuracies above 65%. This suggests that these two classes possess similar characteristics, making them more challenging to differentiate.

Overall, the proposed feature extraction method and subsequent classification demonstrate improved class separation and provide insights into the similarities between different types of jets.

a)                                         b)

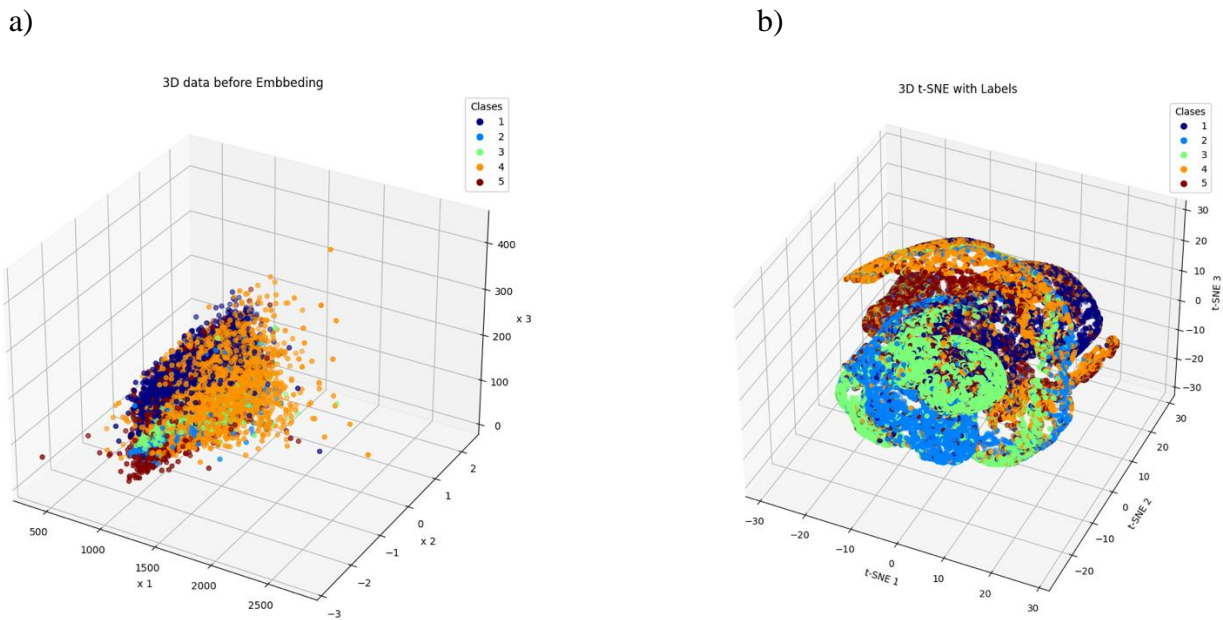

Fig 2. a) data representation of 5 classes. b) data representation after embedding

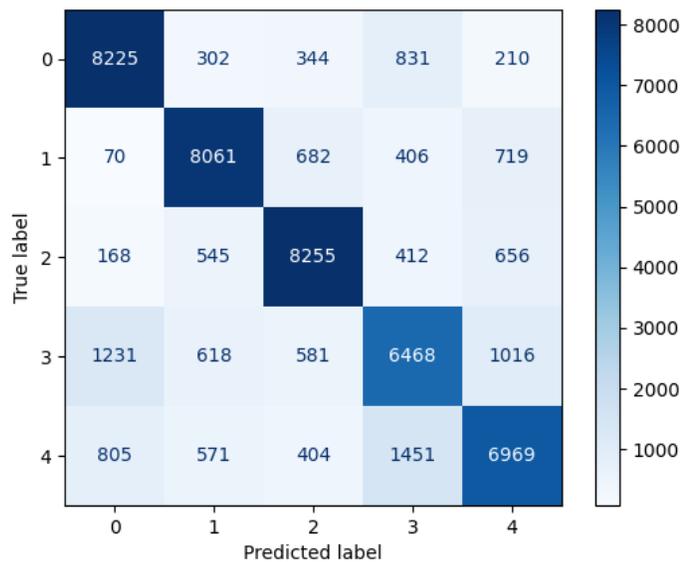

Test accuracy: 0.7598

Accuracy for class 0: 0.8298

Accuracy for class 1: 0.8111

Accuracy for class 2: 0.8225

Accuracy for class 3: 0.6524

Accuracy for class 4: 0.6832

Fig 3. Confusion matrix for 5 classes

a)                                                          b)

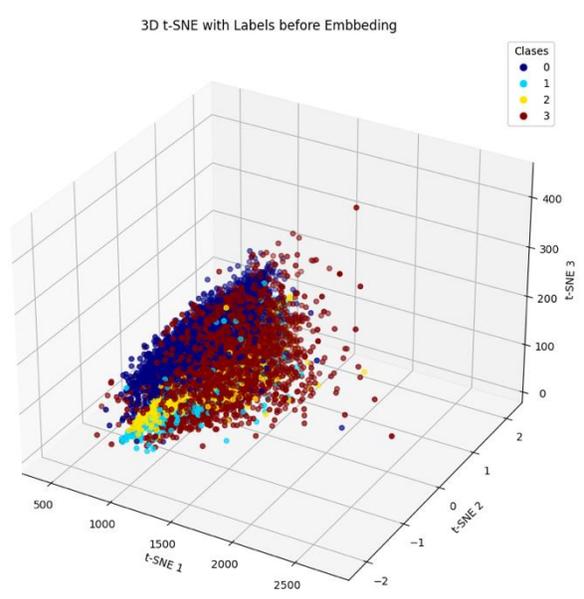 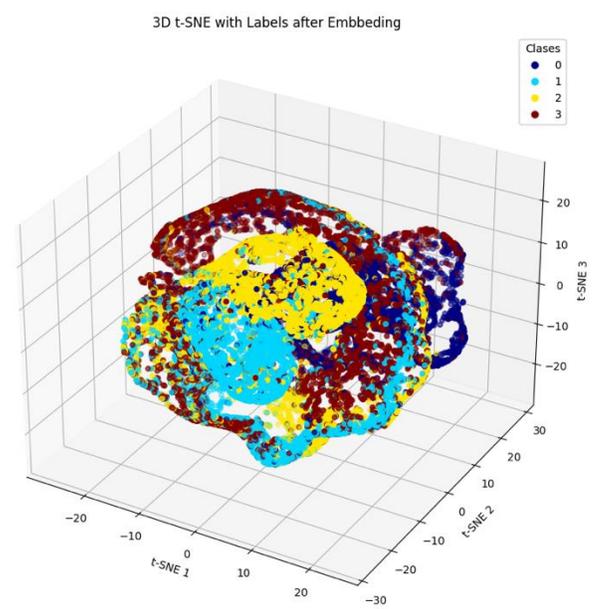

Fig 4. a) data representation of 4 classes TWZQ. b) data representation after embedding

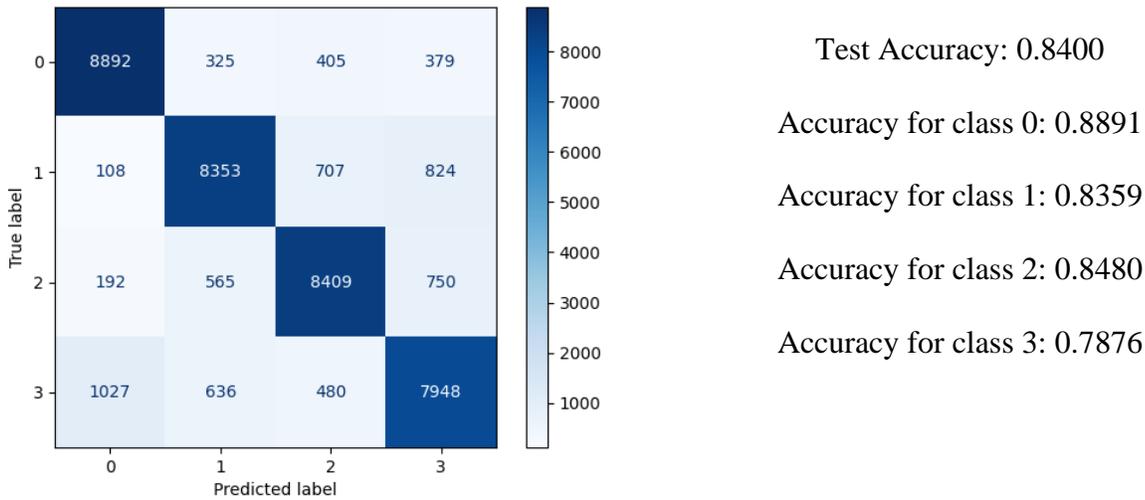

Test Accuracy: 0.8400

Accuracy for class 0: 0.8891

Accuracy for class 1: 0.8359

Accuracy for class 2: 0.8480

Accuracy for class 3: 0.7876

Fig 5. Confusion matrix for 4 classes TWZQ

Figure 4 exhibits a similar pattern as seen in Figure 1, where the data post-embedding demonstrates improved clustering and distribution.

Once the 5 original classes were classified, we proceeded to explore classification using only 4 classes to assess if the overall accuracy improved. It was observed that classes 4 and 5, corresponding to gluons and light quarks, showed lower classification accuracy. Therefore, a configuration with 4 classes was explored, keeping the first 3 classes fixed and combining the fourth class with gluons and light quarks.

The results with light quarks are shown in the confusion matrix in Figure 5. It can be clearly seen that the overall average accuracy increased to 84%, with the class of light quarks achieving 78.7% classification accuracy.

The results with gluons are now presented in the confusion matrix in Figure 6. An overall accuracy of 82.2% was achieved, slightly lower than when light quarks were used as the fourth class, with a lower per-class accuracy of 73.3%. These findings suggest that the class of gluons is more challenging to discriminate, possibly due to similarities with other classes.

These results indicate that the gluon class poses a significant challenge in terms of discrimination, suggesting substantial similarities with other classes in the dataset.

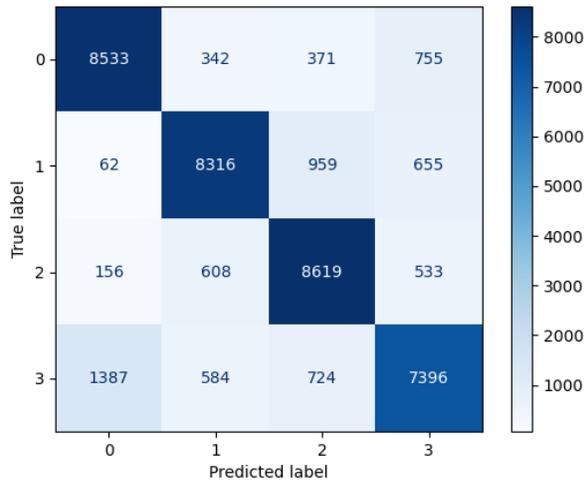

Test Accuracy: 0.8216

Accuracy for class 0: 0.8532

Accuracy for class 1: 0.8322

Accuracy for class 2: 0.8692

Accuracy for class 3: 0.7329

Fig 6. Confusion matrix for 4 classes TWZG

Finally, the classification results for the first 3 classes—top quark, boson, and boson jets—are shown in Figure 7. An increase in overall accuracy to 89.2% is observed, with classification accuracy exceeding 90% for classes 1 and 3. These results indicate that these three types of jets possess distinct characteristics that make them distinguishable from other types of jets.

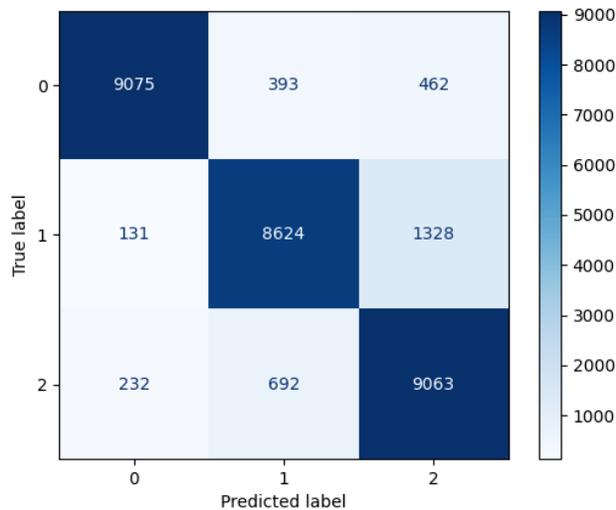

Test Accuracy: 0.8921

Accuracy for class 0: 0.9139

Accuracy for class 1: 0.8553

Accuracy for class 2: 0.9075

Fig 7. Confusion matrix for 4 classes TWZG

## Conclusions

Our study aimed to develop an advanced architecture for jet feature extraction and apply a Deep Neural Network (DNN) for accurate classification, specifically targeting jets from particle collision data. We successfully designed and implemented a machine learning architecture optimized for extracting features from jets, leveraging four-momentum vectors and advanced representation techniques. This approach enabled us to capture comprehensive information about the jets' properties, significantly enhancing our ability to classify them.

The application of our Deep Neural Network (DNN) demonstrated promising results in classifying five distinct jet types: top quark, W boson, Z boson, gluon, and light quark. Initially challenging due to their inherent similarities, the jets were effectively distinguished post-feature extraction. Our method not only improved class separation, as evidenced in the clarity of results shown in Figure 2b and Figure 4, but also provided valuable insights into the nature of jet similarities and discriminability.

Furthermore, our classification results, depicted in the confusion matrices (Figures 3, 5, and 6), underscore the effectiveness of our approach. We achieved over 89% accuracy for the first three classes (top quark, W boson, Z boson), highlighting their distinct and discernible characteristics. However, classes involving gluons and light quarks, despite achieving respectable accuracies, posed greater challenges due to their overlapping features.

In exploring a reduced classification scenario with four classes, we observed improvements in overall accuracy when combining gluons and light quarks into a single class. This adjustment indicated that these two classes share significant characteristics, making their differentiation more complex.

In conclusion, our approach to feature extraction and DNN-based classification significantly advances the accuracy and efficiency of jet classification in particle physics analysis. By enhancing our ability to discern subtle differences among jet types, we contribute to the broader goal of uncovering new physics phenomena from Large Hadron Collider (LHC) data. The achieved results affirm the distinctiveness of top quark, W boson, and Z boson jets, while highlighting ongoing challenges in discriminating gluons and light quarks.

## Acknowledgments
We thank the artificial intelligence imaging group (aiig) laboratory at the University of Puerto Rico, Mayaguez for the computational facility. We thank the authors of JetNet dataset for making it publicly available. This work was supported by US NSF Award 2334265.